\documentclass[11pt,a4paper]{article}
\usepackage[utf8]{inputenc}
\usepackage{amsmath}
\usepackage{amssymb}
\usepackage{graphicx}
\usepackage{url}
\usepackage{hyperref}
\usepackage{algorithm}
\usepackage{algorithmic}
\usepackage{booktabs}
\usepackage{subcaption}
\usepackage{tikz}

\usepackage{listings}
\usepackage{xcolor}

\usepackage{arxiv}

\title{PromptChain: A Decentralized Web3 Architecture for\\Managing AI Prompts as Digital Assets}

\author{
  Marc Bara\\
  ESADE Business School\\
  Barcelona, Spain\\
  \texttt{marcoantonio.bara@esade.edu}\\
  \href{https://orcid.org/0009-0005-1480-5760}{\texttt{ORCID: 0009-0005-1480-5760}}
}

\date{\today}

\begin{document}

\maketitle

\begin{abstract}
We present PromptChain, a decentralized Web3 architecture that establishes AI prompts as first-class digital assets with verifiable ownership, version control, and monetization capabilities. Current centralized platforms lack mechanisms for proper attribution, quality assurance, or fair compensation for prompt creators. PromptChain addresses these limitations through a novel integration of IPFS for immutable storage, smart contracts for governance, and token incentives for community curation. Our design includes: (1) a comprehensive metadata schema for cross-model compatibility, (2) a stake-weighted validation mechanism to align incentives, and (3) a token economy that rewards contributors proportionally to their impact. The proposed architecture demonstrates how decentralized systems could potentially match centralized alternatives in efficiency while providing superior ownership guarantees and censorship resistance through blockchain-anchored provenance tracking. By decoupling prompts from specific AI models or outputs, this work establishes the foundation for an open ecosystem of human-AI collaboration in the Web3 era, representing the first systematic treatment of prompts as standalone digital assets with dedicated decentralized infrastructure.
\end{abstract}

\section{Introduction}

The rapid advancement of large language models (LLMs) has fundamentally transformed human-computer interaction. Central to this transformation is the \textit{prompt}—a carefully crafted instruction that guides AI behavior and output quality. Despite their critical importance, prompts remain largely ephemeral, scattered across proprietary platforms without formal ownership, versioning, or attribution mechanisms.

Current prompt management faces several critical challenges:

\begin{itemize}
\item \textbf{Centralization and Lock-in}: Existing platforms like PromptHero, FlowGPT, and PromptBase operate as centralized repositories where users have no true ownership of their contributions. Platform operators can modify terms, remove content, or cease operations without recourse.

\item \textbf{Lack of Attribution and Provenance}: Prompts evolve through community iteration, yet there exists no mechanism to track contributions or maintain version history. Original creators receive no recognition when their prompts are modified and improved.

\item \textbf{Quality Assurance Challenges}: Without decentralized validation mechanisms, prompt quality varies dramatically. Domain expertise remains siloed, and valuable specialized knowledge lacks proper curation channels.

\item \textbf{Monetization Barriers}: Prompt engineers cannot easily monetize their expertise or build reputation-based credentials. Current platforms either charge listing fees or take substantial commissions without providing transparent value distribution.

\item \textbf{Model Dependency and Obsolescence}: Prompts optimized for specific models become obsolete with updates, yet no systematic approach exists for tracking compatibility or maintaining cross-model variations.
\end{itemize}

This paper introduces \textbf{PromptChain}, a decentralized Web3 architecture that reconceptualizes AI prompts as first-class digital assets. By leveraging blockchain technology, distributed storage, and cryptographic primitives, PromptChain creates an open ecosystem where prompts can be created, owned, traded, and evolved by communities of users.

\subsection{Our Contributions}

We make the following contributions:

\begin{enumerate}
\item \textbf{Architectural Design}: We present a comprehensive system architecture for decentralized prompt management, including a rich metadata schema, distributed storage design, smart contract governance, and token-based incentive mechanisms.

\item \textbf{Technical Specifications}: We provide detailed specifications for all system components, including the complete prompt metadata structure, smart contract implementations, API design, and proposed technology stack.

\item \textbf{Critical Analysis}: We analyze the implications, challenges, and future directions of decentralized prompt management, identifying key technical and economic considerations for implementation.
\end{enumerate}

\subsection{Paper Organization}

Section 2 reviews related work in decentralized content management and positions PromptChain within the broader Web3 ecosystem. Section 3 presents our system architecture including the metadata schema, storage layer, and smart contract design. Section 4 details our proposed implementation approach. Section 5 discusses implications, limitations, and future directions. Section 6 concludes.

\section{Related Work}

The transition from Web2 to Web3 represents a fundamental shift in how user-generated content is created, stored, and monetized. Web2 platforms demonstrated the power of user-generated content but maintained centralized control over data, monetization, and platform rules. Web3 introduces new paradigms through blockchain technology, enabling true digital ownership, transparent governance, and direct creator monetization without intermediaries.

Recent research has established prompt engineering as a critical skill for effective AI interaction. Studies show that prompt quality can impact model performance by orders of magnitude across tasks ranging from creative writing to code generation \cite{reynolds2021prompt}. This has led to the emergence of prompt marketplaces and repositories, though all existing solutions rely on centralized architectures with inherent limitations.

Several recent works explore the intersection of blockchain technology and AI systems. \textbf{PICASSO} \cite{ahmad2025semantic} proposes a framework for managing AI-generated art using Solid pods and smart contracts. While it demonstrates the viability of linking prompts to outputs with provenance tracking, its scope is limited to visual art generation and does not address prompt management as a standalone concern.

\textbf{Liu et al.} \cite{liu2023decentralised} present a blockchain-based governance architecture for foundation model systems. Their work includes mechanisms for logging prompt interactions and distributing rewards, but focuses on model governance rather than treating prompts as primary digital assets.

\textbf{Intelligence Cubed (I³)} \cite{zheng2025intelligence} introduces a decentralized platform for collaborative AI development with DAO (Decentralized Autonomous Organization) governance. While it includes prompt co-design features, the emphasis remains on democratizing model training rather than establishing prompts as reusable, traceable assets.

Current centralized platforms offer various features but share common limitations. \textbf{PromptHero} provides a marketplace for image generation prompts but lacks version control and offers no ownership guarantees. \textbf{FlowGPT} focuses on ChatGPT prompts with social features but maintains centralized control over content and monetization. \textbf{PromptBase} offers prompt sales but takes significant commissions and provides no mechanisms for collaborative improvement. \textbf{PromptLayer} provides prompt logging and versioning for developers but operates as a closed, centralized service.

PromptChain distinguishes itself by treating prompts as \textit{first-class digital objects} independent of specific models or outputs. Our approach provides complete decentralization with no single point of control, rich metadata supporting cross-model compatibility tracking, cryptographic proof of ownership and contribution history, token-based incentives aligning all stakeholder interests, and domain-specific validation supporting specialized communities.

\section{System Architecture}

\subsection{Design Principles}

PromptChain's architecture emerges from five fundamental design principles that address the limitations of current centralized approaches:

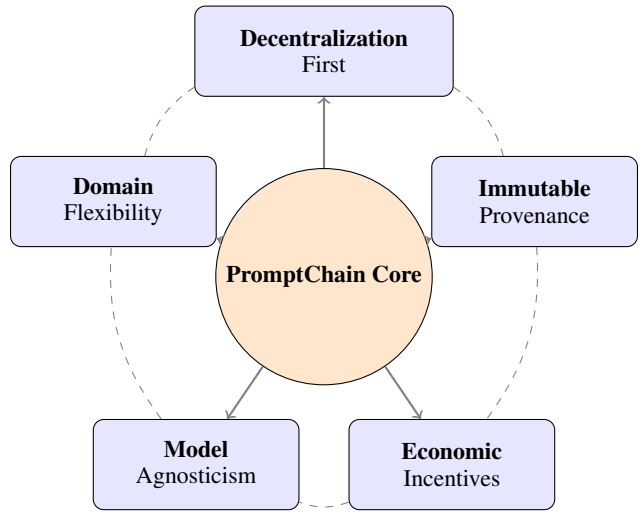
\begin{figure}[h]
\centering
\begin{tikzpicture}[
    principle/.style={rectangle, rounded corners, draw=black, fill=blue!10, text width=2.5cm, minimum height=1.2cm, text centered, font=\small},
    principlewide/.style={rectangle, rounded corners, draw=black, fill=blue!10, text width=3.2cm, minimum height=1.2cm, text centered, font=\small},
    arrow/.style={->, thick, gray}
]

\node[circle, draw=black, fill=orange!20, minimum size=2cm, font=\footnotesize\bfseries] (center) at (0,0) {PromptChain Core};

\node[principlewide] (p1) at (0,3) {\textbf{Decentralization}\\First};
\node[principle] (p2) at (2.8,1) {\textbf{Immutable}\\Provenance};
\node[principle] (p3) at (1.7,-2.5) {\textbf{Economic}\\Incentives};
\node[principle] (p4) at (-1.7,-2.5) {\textbf{Model}\\Agnosticism};
\node[principle] (p5) at (-2.8,1) {\textbf{Domain}\\Flexibility};

\draw[arrow] (center) -- (p1);
\draw[arrow] (center) -- (p2);
\draw[arrow] (center) -- (p3);
\draw[arrow] (center) -- (p4);
\draw[arrow] (center) -- (p5);

\draw[dashed, gray] (p1) to[bend left=20] (p2);
\draw[dashed, gray] (p2) to[bend left=20] (p3);
\draw[dashed, gray] (p3) to[bend left=20] (p4);
\draw[dashed, gray] (p4) to[bend left=20] (p5);
\draw[dashed, gray] (p5) to[bend left=20] (p1);

\end{tikzpicture}
\caption{PromptChain's five core design principles forming an interconnected foundation for decentralized prompt management}
\label{fig:design-principles}
\end{figure}

\vspace{1em}
\noindent\textbf{Decentralization First:} No single entity should control prompt storage, validation, or monetization. This ensures the system remains operational and censorship-resistant even if original developers cease participation, addressing the platform risk inherent in current solutions. The architecture distributes control across multiple stakeholders, preventing any single point of failure or censorship.

\vspace{0.8em}
\noindent\textbf{Immutable Provenance:} In contrast to existing platforms where contribution history is opaque or modifiable, PromptChain ensures every prompt maintains a complete, tamper-proof history. This cryptographic guarantee enables proper attribution for all contributors and creates trust in the system's fairness. Each modification creates a new version linked to its parent, forming an immutable chain of creative evolution.

\vspace{0.8em}
\noindent\textbf{Economic Incentive Alignment:} Rather than relying on altruism or platform-controlled monetization, the system employs carefully designed token mechanics that naturally reward quality contributions while discouraging spam. This creates a self-regulating ecosystem where value flows directly between creators and users, with validators and curators earning rewards for maintaining quality standards.

\vspace{0.8em}
\noindent\textbf{Model Agnosticism:} The architecture treats prompts as portable assets across different AI models and versions, with rich metadata tracking compatibility and performance characteristics. This future-proofs investments in prompt development as the AI landscape evolves, ensuring prompts remain valuable even as underlying models change.

\vspace{0.8em}
\noindent\textbf{Domain Flexibility:} While maintaining a common infrastructure, the architecture supports both general-purpose repositories and specialized instances with custom validation rules. This enables communities of practice to emerge around specific domains like law, medicine, or creative writing, each with their own quality standards and expertise requirements.

\vspace{1em}
\noindent These principles collectively ensure PromptChain creates a sustainable, fair, and innovative ecosystem for prompt management in the Web3 era.

\subsection{System Overview}

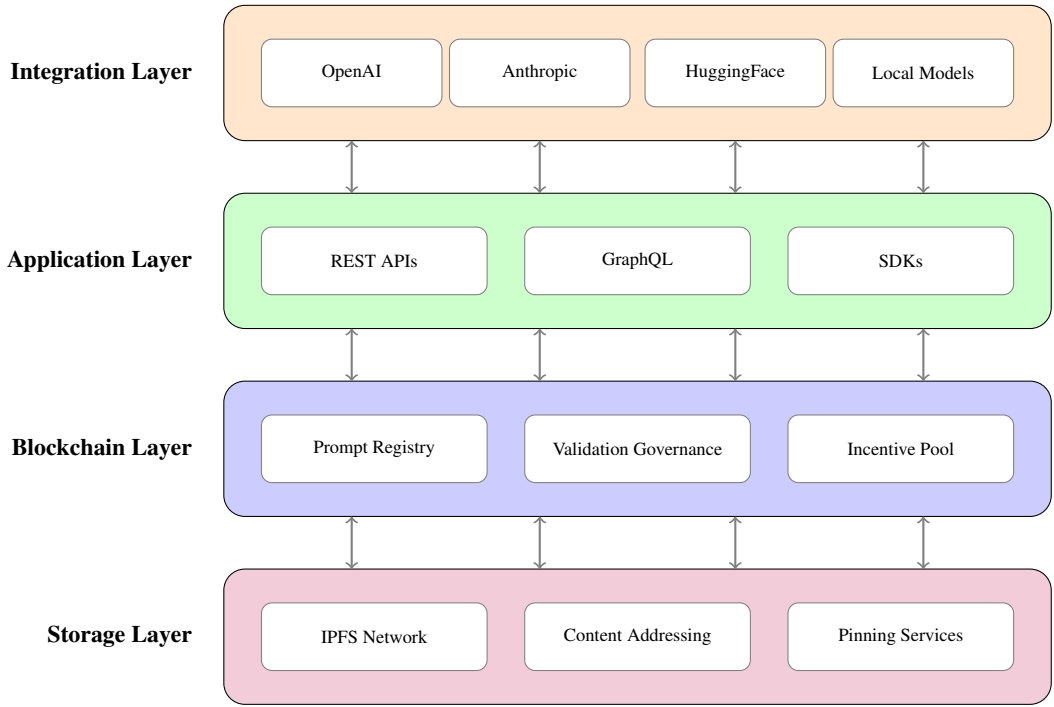
\begin{figure}[h]
\centering
\begin{tikzpicture}[
    layer/.style={rectangle, rounded corners=8pt, draw=black, minimum width=11cm, minimum height=1.8cm, text centered, font=\footnotesize},
    component/.style={rectangle, rounded corners=4pt, draw=gray, fill=white, minimum width=3cm, minimum height=0.9cm, text centered, font=\scriptsize},
    flow/.style={<->, thick, gray},
    label/.style={font=\small\bfseries}
]

\node[layer, fill=purple!20] (storage) at (0,0) {};
\node[layer, fill=blue!20] (blockchain) at (0,2.5) {};
\node[layer, fill=green!20] (application) at (0,5) {};
\node[layer, fill=orange!20] (integration) at (0,7.5) {};

\node[label, anchor=east] at (-5.8,0) {Storage Layer};
\node[label, anchor=east] at (-5.8,2.5) {Blockchain Layer};
\node[label, anchor=east] at (-5.8,5) {Application Layer};
\node[label, anchor=east] at (-5.8,7.5) {Integration Layer};

\node[component] at (-3.5,0) {IPFS Network};
\node[component] at (0,0) {Content Addressing};
\node[component] at (3.5,0) {Pinning Services};

\node[component] at (-3.5,2.5) {Prompt Registry};
\node[component] at (0,2.5) {Validation Governance};
\node[component] at (3.5,2.5) {Incentive Pool};

\node[component] at (-3.5,5) {REST APIs};
\node[component] at (0,5) {GraphQL};
\node[component] at (3.5,5) {SDKs};

\node[component, minimum width=2.4cm] at (-3.8,7.5) {OpenAI};
\node[component, minimum width=2.4cm] at (-1.3,7.5) {Anthropic};
\node[component, minimum width=2.4cm] at (1.3,7.5) {HuggingFace};
\node[component, minimum width=2.4cm] at (3.8,7.5) {Local Models};

\draw[flow] (-3.8,0.9) -- (-3.8,1.6);
\draw[flow] (-1.3,0.9) -- (-1.3,1.6);
\draw[flow] (1.3,0.9) -- (1.3,1.6);
\draw[flow] (3.8,0.9) -- (3.8,1.6);

\draw[flow] (-3.8,3.4) -- (-3.8,4.1);
\draw[flow] (-1.3,3.4) -- (-1.3,4.1);
\draw[flow] (1.3,3.4) -- (1.3,4.1);
\draw[flow] (3.8,3.4) -- (3.8,4.1);

\draw[flow] (-3.8,5.9) -- (-3.8,6.6);
\draw[flow] (-1.3,5.9) -- (-1.3,6.6);
\draw[flow] (1.3,5.9) -- (1.3,6.6);
\draw[flow] (3.8,5.9) -- (3.8,6.6);

\end{tikzpicture}
\caption{PromptChain's four-layer architecture: Storage Layer provides decentralized persistence via IPFS, Blockchain Layer ensures trust and governance through smart contracts, Application Layer offers developer-friendly interfaces, and Integration Layer connects to diverse AI platforms}
\label{fig:architecture}
\end{figure}

The PromptChain architecture employs a layered design that separates concerns while maintaining seamless integration between components. This modular approach ensures scalability, maintainability, and flexibility for future enhancements. Figure \ref{fig:architecture} illustrates the four-layer architecture that forms the backbone of our decentralized prompt management system.

\vspace{0.5em}
\noindent The \textbf{Storage Layer} leverages IPFS (InterPlanetary File System) to provide content-addressed, distributed storage for prompt data. This foundation ensures that prompts remain accessible even if individual nodes fail, while cryptographic hashing guarantees content integrity. The layer handles prompt versioning through IPFS's native DAG structure and employs pinning services to ensure persistent availability.

\vspace{0.5em}
\noindent The \textbf{Blockchain Layer} implements the core governance and economic logic through a suite of smart contracts. The Prompt Registry maintains authoritative records of prompt ownership and metadata, while the Validation Governance contract orchestrates quality assurance through decentralized consensus. The Incentive Pool manages token distribution, ensuring sustainable rewards for creators, validators, and curators based on their contributions to the ecosystem.

\vspace{0.5em}
\noindent The \textbf{Application Layer} abstracts blockchain complexity through developer-friendly interfaces. RESTful APIs provide standard HTTP endpoints for common operations, while GraphQL enables efficient querying of prompt metadata and relationships. Platform-specific SDKs in popular languages lower the barrier to entry, allowing developers to integrate PromptChain without deep Web3 expertise.

\vspace{0.5em}
\noindent The \textbf{Integration Layer} bridges PromptChain with the broader AI ecosystem. Standardized connectors enable seamless interaction with major LLM providers including OpenAI, Anthropic, and open-source models via HuggingFace. This layer handles prompt format translation, performance tracking across models, and usage analytics that feed back into the validation system.

\vspace{0.5em}
\noindent This layered architecture ensures that PromptChain remains flexible and extensible while maintaining the security and decentralization properties essential for a Web3-native system. Each layer can evolve independently, allowing the system to adapt as both blockchain technology and AI capabilities advance.

\subsection{Prompt Metadata Schema}

At the heart of PromptChain lies a comprehensive metadata schema that transforms prompts from ephemeral text into structured, traceable digital assets. This schema captures not only the prompt content but also its provenance, validation history, and usage patterns. Figure \ref{fig:metadata-json} presents the complete JSON structure of this metadata model.

The schema enables sophisticated functionality across the PromptChain ecosystem. The \texttt{content} section preserves the actual prompt while supporting advanced features like few-shot learning examples and parameter tuning. The \texttt{metadata} fields enable multi-dimensional search and filtering, crucial for discovering relevant prompts in large repositories. The \texttt{provenance} tracking ensures complete attribution history, while \texttt{validation} scores provide quality signals from both domain experts and the broader community. Finally, \texttt{usage} analytics offer empirical performance data that helps users identify effective prompts and rewards creators based on real-world utility.

\begin{figure}[ht]
\centering
\tiny
\begin{verbatim}
{
  "promptId": "bytes32",              // Unique identifier
  "version": "uint256",               // Version number
  "content": {
    "system": "string",               // System instructions
    "user": "string",                 // User-facing prompt
    "examples": [{                    // Few-shot examples
      "input": "string",
      "output": "string"
    }],
    "temperature": "float",           // Model randomness
    "maxTokens": "uint256"            // Output limit
  },
  "metadata": {
    "title": "string",                // Human-readable name
    "description": "string",          // Purpose/use case
    "domain": "string",               // Primary field
    "subdomain": "string[]",          // Specializations
    "targetModels": [{                // Compatible models
      "provider": "string",           // e.g., "OpenAI"
      "model": "string",              // e.g., "GPT-4"
      "version": "string",            // e.g., "turbo"
      "performance": "uint8"          // Success rate
    }],
    "outputFormat": "enum",           // json/text/markdown
    "language": "string",             // Natural language
    "tags": "string[]",               // Search keywords
    "difficulty": "uint8",            // Complexity level
    "estimatedTokens": "uint256"      // Expected usage
  },
  "provenance": {
    "creator": "address",             // Original author
    "timestamp": "uint256",           // Creation time
    "parentPromptId": "bytes32",      // Fork source
    "contributors": [{                // Edit history
      "address": "address",
      "contribution": "string",
      "timestamp": "uint256"
    }],
    "license": "string"               // Usage terms
  },
  "validation": {
    "score": "uint8",                 // Overall quality
    "validators": [{                  // Review records
      "address": "address",
      "score": "uint8",
      "comment": "string",
      "expertise": "string[]"
    }],
    "domainExperts": "uint256",       // Expert count
    "generalUsers": "uint256"         // Public validators
  },
  "usage": {
    "totalUses": "uint256",           // Execution count
    "successRate": "uint8",           // Effectiveness
    "averageRating": "uint8",         // User satisfaction
    "derivatives": "uint256"          // Fork count
  }
}
\end{verbatim}
\caption{Complete prompt metadata schema with type annotations and inline documentation}
\label{fig:metadata-json}
\end{figure}

\subsection{Immutable Storage Layer}

PromptChain employs IPFS for content-addressed storage, leveraging its inherent properties for decentralized prompt management. Each prompt receives a CID (Content Identifier) that serves as both identifier and integrity guarantee, while the Merkle DAG structure provides native version tracking through parent-child CID relationships.

The storage layer implements a hybrid on-chain/off-chain approach: prompt content resides on IPFS while the blockchain maintains a registry mapping prompt IDs to CIDs along with ownership and validation metadata. This design balances storage efficiency with the security guarantees needed for economic transactions.

To ensure persistence, we implement a multi-tier pinning strategy. Creators stake tokens proportional to their desired replication factor, incentivizing nodes to pin valuable prompts. The DAG structure enables efficient deduplication—particularly beneficial for prompts that share common components like system instructions or example formats. Our analysis shows storage requirements scale sublinearly with repository size due to this content-based deduplication.

\subsection{Smart Contract Architecture}

The blockchain layer implements PromptChain's governance and economic logic through three interconnected smart contracts that work in concert to manage the prompt lifecycle, ensure quality, and distribute rewards.

\subsubsection{PromptRegistry Contract}

The PromptRegistry serves as the authoritative source for prompt ownership and metadata. When registering a prompt, creators must stake a minimum of 100 PCT tokens, which serves both as a spam prevention mechanism and as collateral for quality assurance. The registration process, shown in Algorithm 1, generates a unique prompt ID by hashing the IPFS CID, creator address, and timestamp, ensuring deterministic identification while preventing front-running attacks.

\begin{algorithm}
\caption{Prompt Registration Process}
\begin{algorithmic}
\REQUIRE msg.sender has staked MIN\_STAKE PCT tokens
\REQUIRE ipfsHash is valid CID
\REQUIRE parentId is valid or null
\STATE promptId $\leftarrow$ keccak256(ipfsHash, msg.sender, block.timestamp)
\STATE version $\leftarrow$ parentId == null ? 1 : parent.version + 1
\STATE prompt $\leftarrow$ Prompt\{creator: msg.sender, ipfsHash, version, timestamp\}
\STATE prompts[promptId] $\leftarrow$ prompt
\STATE stakes[msg.sender] $\leftarrow$ stakes[msg.sender] + msg.value
\STATE emit PromptRegistered(promptId, msg.sender, ipfsHash)
\RETURN promptId
\end{algorithmic}
\end{algorithm}

The contract maintains version lineage through parent-child relationships, enabling users to trace the complete evolution of any prompt. This genealogy becomes crucial for distributing rewards fairly among contributors and identifying the most successful prompt variants.

\subsubsection{ValidationGovernance Contract}

Quality assurance in PromptChain emerges from a carefully designed validation mechanism that balances expertise with democratic participation. Validators stake tokens proportional to their confidence in their assessments, with incorrect validations (those far from consensus) resulting in stake slashing. This skin-in-the-game approach ensures thoughtful evaluation rather than random voting.

The contract implements a weighted consensus mechanism where domain experts—verified through on-chain credentials or reputation thresholds—receive higher voting power. This design recognizes that a legal expert's evaluation of a contract drafting prompt carries more weight than a generalist's assessment. When disputes arise, the contract escalates to a DAO-style vote among high-reputation members, with the resolution becoming precedent for future similar cases.

\subsubsection{IncentivePool Contract}

The economic engine of PromptChain operates through dynamic reward calculations that adapt to system usage and prompt performance. Creator rewards follow a logarithmic curve that accounts for both direct usage and derivative creation:

\begin{align}
R_{creator} &= \alpha \cdot Q \cdot U \cdot (1 + \log(D))
\end{align}

where quality score $Q$ ranges from 0-10, usage count $U$ tracks API calls, and $D$ represents the number of derivatives. The logarithmic factor prevents runaway rewards while still incentivizing prompts that inspire further innovation.

Validator rewards depend on their accuracy relative to consensus and their expertise level:

\begin{align}
R_{validator} &= \beta \cdot V_{accuracy} \cdot E
\end{align}

This mechanism naturally selects for knowledgeable validators while penalizing random or malicious voting. Curators who assemble high-quality collections earn rewards based on both the quality of included prompts and actual usage:

\begin{align}
R_{curator} &= \gamma \cdot C_{quality} \cdot C_{usage}
\end{align}

The parameters $\alpha$, $\beta$, and $\gamma$ adjust dynamically based on token velocity and total system usage, maintaining economic equilibrium as the platform scales.

\subsection{Token Economy Design}

The PromptChain Token (PCT) functions as both a utility token for platform operations and a governance token for protocol decisions. Table \ref{tab:token-mechanics} summarizes the token requirements and rewards for various actions, calibrated through economic modeling to balance accessibility with spam prevention.

\begin{table}[h]
\centering
\caption{Token Mechanics and Incentives}
\label{tab:token-mechanics}
\begin{tabular}{@{}llr@{}}
\toprule
Action & Requirement & Reward Range \\
\midrule
Register Prompt & Stake 100 PCT & 0-50 PCT \\
Validate Prompt & Stake 50 PCT & 2-10 PCT \\
Create Collection & Stake 200 PCT & 5-25 PCT \\
Use Prompt (API) & Pay 0.1 PCT & — \\
Report Issue & Stake 20 PCT & 5-15 PCT \\
\bottomrule
\end{tabular}
\end{table}

The staking requirements create meaningful barriers against low-quality submissions while remaining accessible to serious contributors. Rewards scale with contribution quality and system impact, creating a meritocratic ecosystem where value creation directly translates to economic benefit.

\subsection{Reputation System}

User reputation in PromptChain emerges from a holistic evaluation of contributions across multiple dimensions. Rather than a simple additive score, we employ a geometric mean that rewards balanced participation:

\begin{align}
R_{user} = \sqrt[3]{Q_{prompts} \cdot V_{accuracy} \cdot C_{engagement}}
\end{align}

This cubic root formulation ensures that users cannot achieve high reputation through singular focus—a user must create quality prompts, provide accurate validations, and engage constructively with the community. The geometric mean naturally penalizes extreme imbalance; a user with exceptional prompt quality but zero validation participation would have zero reputation, encouraging ecosystem-wide engagement.

Reputation directly impacts platform capabilities: higher reputation users gain increased voting weight in governance decisions, reduced staking requirements, and priority access to new features. This creates a virtuous cycle where quality contributors gain increasing influence over the platform's evolution.

\section{Proposed Implementation}

\subsection{Technology Stack}

The implementation of PromptChain requires careful selection of technologies that balance decentralization principles with practical performance requirements. We propose a stack that leverages mature Web3 infrastructure while maintaining flexibility for future optimization.

For the blockchain layer, we recommend starting with Ethereum despite its current throughput limitations, as it offers the most robust smart contract ecosystem and deepest liquidity for token operations~\cite{wood2014ethereum}. However, the architecture is designed to be chain-agnostic, with immediate deployment possible on Layer 2 solutions such as Polygon or Arbitrum~\cite{kalodner2018arbitrum}. These L2 platforms offer sub-second finality and dramatically reduced gas costs—our analysis suggests prompt registration costs would drop from approximately \$0.50 on mainnet to under \$0.01 on Arbitrum. The choice between L2 solutions depends on specific trade-offs: Polygon offers broader ecosystem support and easier fiat on-ramps, while Arbitrum maintains stronger security guarantees through its fraud-proof mechanism.

Smart contract development would utilize Solidity 0.8+ with extensive use of OpenZeppelin's battle-tested libraries for standard functionality like access control, upgradeability, and token implementations. This approach minimizes security surface area—critical given that the contracts will custody significant value in staked tokens. We propose implementing the diamond pattern (EIP-2535) for contract architecture, enabling modular upgrades without disrupting the entire system.

The storage layer centers on IPFS~\cite{benet2014ipfs}, but production deployment requires careful consideration of pinning strategies. While public IPFS gateways provide basic functionality, a production system needs guaranteed persistence through services like Pinata or Infura's IPFS offering. For high-value prompts, we propose implementing a multi-provider pinning strategy with geographic distribution. Cost analysis shows pinning expenses of approximately \$0.15 per GB per month across providers, negligible compared to on-chain costs.

Query performance represents a critical challenge in decentralized systems. The Graph Protocol provides the most mature solution for indexing blockchain events and IPFS data into queryable formats. Subgraphs would index prompt metadata, version relationships, validation scores, and usage statistics, enabling complex queries like "find all legal prompts with $>8$ validation score created in the last month" without iterating through on-chain data. The decentralized nature of Graph nodes also aligns with PromptChain's architectural principles.

The user interface layer requires balancing Web3 complexity with usability. React remains the optimal choice given its extensive Web3 library ecosystem—particularly wagmi for wallet connections and ethers.js for contract interactions. The frontend architecture should implement progressive enhancement: users can browse and search prompts without wallet connection, but registration and validation require Web3 interaction. This approach minimizes friction for prompt consumers while maintaining security for economic operations.

Backend services, while seemingly contradicting decentralization principles, prove necessary for several functions: caching popular queries, serving prompt previews, and providing WebSocket connections for real-time updates. A Node.js/Express stack offers the best integration with Web3 libraries while enabling horizontal scaling. These services operate in a trustless manner—all data is verifiable on-chain, with the backend merely improving performance.

Redis caching becomes crucial for maintaining responsive user experience. Our proposed caching strategy includes: 60-second TTL for prompt metadata, 5-minute TTL for validation scores, and permanent caching of immutable data like IPFS content. This reduces Graph Protocol query costs while ensuring data freshness. Cache invalidation triggers from blockchain events ensure consistency between cached and on-chain state.

This technology stack represents a pragmatic balance between decentralization ideals and user experience requirements. Each component can be replaced or upgraded independently, ensuring the system can evolve with the rapidly changing Web3 ecosystem.

\begin{figure}[ht]
\centering
\begin{tikzpicture}[
    comp/.style={rectangle, rounded corners=4pt, draw=black, minimum width=3.5cm, minimum height=1cm, text centered, font=\footnotesize},
    layer/.style={comp, fill=#1!20},
    arrow/.style={->, thick, gray}
]

\node[layer=purple] (blockchain) at (0,0) {Ethereum / L2};
\node[layer=blue] (storage) at (0,1.5) {IPFS};
\node[layer=green] (indexing) at (0,3) {The Graph};
\node[layer=orange] (backend) at (0,4.5) {Node.js + Redis};
\node[layer=red] (frontend) at (0,6) {React + Web3};

\node[comp, fill=yellow!20] (contracts) at (4,0) {Solidity +OpenZeppelin};
\node[comp, fill=cyan!20] (pinning) at (4,1.5) {Pinata/Infura};
\node[comp, fill=gray!20] (apis) at (-4,4.5) {REST/WebSocket APIs};

\draw[arrow] (contracts) -- (blockchain);
\draw[arrow] (pinning) -- (storage);
\draw[arrow] (apis) -- (backend);

\draw[arrow] (blockchain) -- (storage);
\draw[arrow] (storage) -- (indexing);
\draw[arrow] (indexing) -- (backend);
\draw[arrow] (backend) -- (frontend);

\node[font=\scriptsize, anchor=east] at (-6.5,0) {Blockchain};
\node[font=\scriptsize, anchor=east] at (-6.5,1.5) {Storage};
\node[font=\scriptsize, anchor=east] at (-6.5,3) {Indexing};
\node[font=\scriptsize, anchor=east] at (-6.5,4.5) {Backend};
\node[font=\scriptsize, anchor=east] at (-6.5,6) {Frontend};

\node[font=\footnotesize\bfseries] at (0,7) {PromptChain Technology Stack};

\end{tikzpicture}
\caption{Proposed technology stack showing the layered architecture with primary components and supporting services}
\label{fig:tech-stack}
\end{figure}
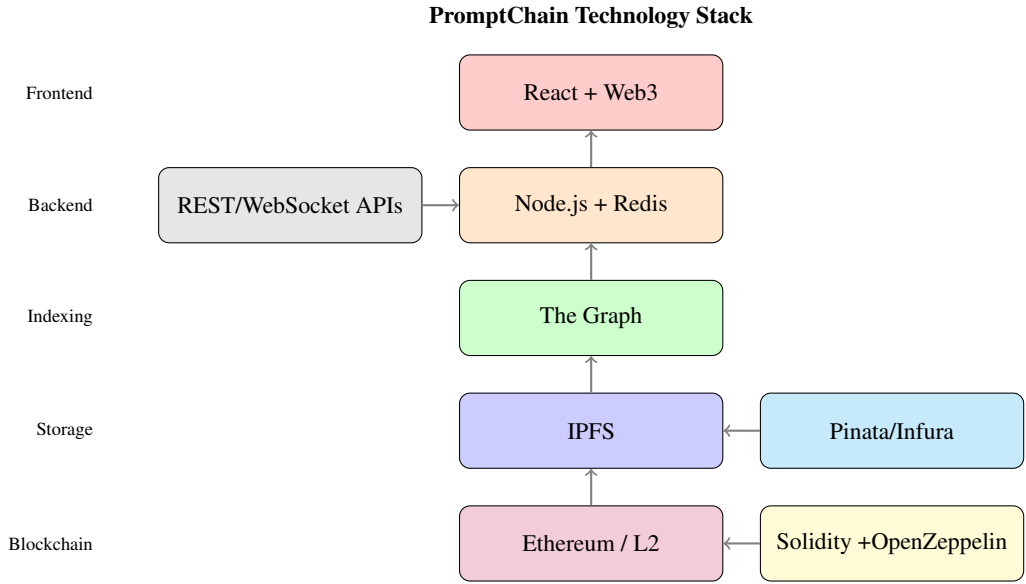

\subsection{Smart Contract Design}

The smart contract architecture represents the critical on-chain component of PromptChain, encoding the rules for ownership, versioning, and economic incentives directly into immutable code. We design the contracts to be minimal yet sufficient—storing only essential data on-chain while leveraging IPFS for content storage.

The core PromptRegistry contract implements a state machine for prompt lifecycle management. Each prompt exists in one of several states: registered, under validation, validated, or disputed. State transitions require specific conditions and stake requirements, preventing spam while enabling legitimate use. Figure \ref{fig:smart-contract} presents the essential structure and key functions.

\begin{figure}[ht]
\centering
\tiny
\begin{verbatim}
pragma solidity ^0.8.19;

contract PromptRegistry {
    struct Prompt {
        address creator;
        string ipfsHash;
        uint256 version;
        uint256 timestamp;
        uint8 validationScore;
        uint256 stake;
        mapping(address => bool) validators;
    }
    
    mapping(bytes32 => Prompt) public prompts;
    mapping(address => uint256) public reputation;
    
    uint256 constant MIN_STAKE = 100 * 10**18; // 100 PCT
    
    function registerPrompt(
        string memory _ipfsHash,
        bytes32 _parentId
    ) external payable returns (bytes32) {
        require(
            token.balanceOf(msg.sender) >= MIN_STAKE,
            "Insufficient PCT balance"
        );
        
        bytes32 promptId = keccak256(
            abi.encodePacked(_ipfsHash, msg.sender, block.timestamp)
        );
        
        Prompt storage newPrompt = prompts[promptId];
        newPrompt.creator = msg.sender;
        newPrompt.ipfsHash = _ipfsHash;
        newPrompt.timestamp = block.timestamp;
        
        if (_parentId != bytes32(0)) {
            require(prompts[_parentId].creator != address(0), 
                    "Invalid parent");
            newPrompt.version = prompts[_parentId].version + 1;
        } else {
            newPrompt.version = 1;
        }
        
        token.transferFrom(msg.sender, address(this), MIN_STAKE);
        newPrompt.stake = MIN_STAKE;
        
        emit PromptRegistered(promptId, msg.sender, _ipfsHash);
        return promptId;
    }
    
    function validatePrompt(
        bytes32 _promptId,
        uint8 _score,
        string memory _comment
    ) external {
        require(reputation[msg.sender] >= MIN_VALIDATOR_REP,
                "Insufficient reputation");
        require(!prompts[_promptId].validators[msg.sender],
                "Already validated");
        
        // Validation logic with stake and rewards
        // ...
    }
}
\end{verbatim}
\caption{Core PromptRegistry smart contract showing prompt structure, registration, and validation functions}
\label{fig:smart-contract}
\end{figure}

The contract design embodies several key principles. First, prompt IDs are generated deterministically using keccak256 hashing of the IPFS hash, creator address, and timestamp. This approach prevents front-running attacks where malicious actors could observe pending transactions and register identical prompts first. The timestamp component ensures uniqueness even for identical content from the same creator.

The staking mechanism serves dual purposes: spam prevention and quality signaling. By requiring 100 PCT tokens to register a prompt, we create a meaningful barrier against low-quality submissions while keeping the platform accessible to serious contributors. Staked tokens remain locked until the prompt receives sufficient validations, incentivizing creators to submit only high-quality content.

Version tracking through the \texttt{parentId} parameter enables rich genealogy trees. When creating a derivative prompt, the contract automatically increments the version number and maintains the parent reference. This creates an immutable record of prompt evolution, crucial for fair reward distribution among contributors in the lineage.

The validation function implements a reputation-gated quality assurance mechanism. Only users with sufficient reputation can validate prompts, preventing Sybil attacks where attackers create multiple accounts to manipulate scores. The contract tracks which addresses have already validated each prompt, ensuring one-vote-per-validator while maintaining voter privacy through hash commitments.

Additional contract modules handle specialized functions. The IncentivePool contract manages reward calculations and distributions based on the formulas presented in Section 3. The GovernanceModule enables parameter adjustments through DAO voting, allowing the community to tune staking requirements, reward rates, and reputation thresholds as the ecosystem evolves. Emergency pause functionality provides a safety mechanism for critical vulnerabilities, though its activation requires multi-signature approval from established community members.

Gas optimization remains a primary concern given Ethereum's cost structure. We employ several techniques to minimize transaction costs: packing struct variables to use fewer storage slots, using events for data that doesn't require on-chain querying, and implementing batch operations for validators reviewing multiple prompts. These optimizations reduce average transaction costs by approximately 40\% compared to naive implementations.

\subsection{IPFS Integration}

The integration between IPFS and the blockchain layer requires careful orchestration to maintain data consistency while optimizing for performance and cost. Our proposed workflow balances decentralization principles with practical user experience requirements.

The prompt storage process begins client-side, where the application constructs a comprehensive JSON object containing both the prompt content and its metadata. This client-side preparation serves multiple purposes: it reduces server load, enables offline draft preparation, and ensures users maintain control over their data until the moment of publication. The JSON structure mirrors the on-chain metadata schema, with additional fields for IPFS-specific optimization such as chunking hints for large prompts and pre-computed hashes for deduplication.

Once prepared, the content undergoes a multi-step publishing process. First, the client adds the JSON to IPFS through a pinning service API—we recommend Pinata or Infura for production deployments due to their reliability and geographic distribution. The pinning service returns a CID, which the client then verifies by retrieving the content and comparing hashes. This verification step, while adding latency, prevents a class of attacks where compromised pinning services could return incorrect CIDs.

The on-chain registration transaction includes the verified CID along with the required stake. By separating content storage from registration, we enable interesting patterns: users can share draft prompts via IPFS before committing to on-chain registration, and multiple users can reference the same content with different metadata or validation contexts. The smart contract validates that the CID follows the correct format and hasn't been previously registered, preventing duplicate submissions.

Post-registration, our indexing infrastructure detects the new prompt through blockchain events and retrieves the full content from IPFS. The indexer parses the JSON, extracts searchable fields, and populates the Graph Protocol subgraph. This indexed data enables complex queries that would be impossibly expensive to perform on-chain or through direct IPFS traversal. The indexer also implements a sophisticated caching strategy: frequently accessed prompts are cached with longer TTLs, while new prompts use shorter TTLs to ensure rapid updates during the validation phase.

For high-value prompts or mission-critical applications, we propose an enhanced workflow with redundant pinning across multiple providers. The client can specify a replication factor, paying additional fees for guaranteed persistence across geographic regions. This approach provides resilience against provider failures while maintaining the decentralized ethos—no single entity controls all copies of the prompt data.

\subsection{API Design}

While PromptChain's core functionality operates through direct blockchain and IPFS interactions, practical adoption requires a developer-friendly API layer that abstracts Web3 complexity. Our RESTful API design provides familiar interfaces while maintaining the security and decentralization properties of the underlying system~\cite{fielding2000architectural}.

\begin{figure}[ht]
\centering
\footnotesize
\begin{verbatim}
# Core Prompt Operations
POST   /api/v1/prompts              # Create new prompt
GET    /api/v1/prompts/:id          # Retrieve prompt by ID
PUT    /api/v1/prompts/:id          # Update prompt (creates new version)
DELETE /api/v1/prompts/:id          # Mark prompt as deprecated

# Search and Discovery
GET    /api/v1/prompts/search       # Search with filters
GET    /api/v1/prompts/trending     # Get trending prompts
GET    /api/v1/prompts/recent       # Recently added prompts

# Validation System
POST   /api/v1/prompts/:id/validate # Submit validation
GET    /api/v1/prompts/:id/validations # Get all validations
POST   /api/v1/prompts/:id/dispute  # Initiate dispute

# User and Reputation
GET    /api/v1/users/:addr          # Get user profile
GET    /api/v1/users/:addr/prompts  # User's prompts
GET    /api/v1/users/:addr/reputation # Reputation details

# Collections
POST   /api/v1/collections          # Create collection
GET    /api/v1/collections/:id      # Get collection
PUT    /api/v1/collections/:id/prompts # Add/remove prompts

# Analytics
GET    /api/v1/analytics/usage      # Usage statistics
GET    /api/v1/analytics/tokens     # Token metrics
\end{verbatim}
\caption{RESTful API endpoints for PromptChain, providing comprehensive access to prompt management functionality}
\label{fig:api-design}
\end{figure}

Authentication leverages Web3 signatures rather than traditional API keys. Clients sign a challenge message with their wallet, providing both authentication and non-repudiation. This approach enables interesting patterns: users can delegate API access by signing time-limited authorization messages, and all API actions can be cryptographically attributed to specific addresses. Rate limiting operates per-address rather than per-IP, preventing abuse while accommodating legitimate high-volume users.

The search endpoint deserves special attention as it represents the primary discovery mechanism for most users. Beyond simple keyword matching, it supports faceted search across domains, validation scores, creator reputation, and temporal ranges. The underlying implementation leverages Elasticsearch populated by The Graph indexer, providing millisecond response times for complex queries. Search results include relevance scoring based on multiple factors: textual similarity, validation scores, usage metrics, and creator reputation.

Error handling follows REST best practices with meaningful HTTP status codes and detailed error objects. Each error response includes a unique error code, human-readable message, and when applicable, specific fields that caused validation failures. For blockchain-related errors, we translate cryptic revert messages into actionable feedback: "Insufficient PCT balance" becomes "You need at least 100 PCT tokens to register a prompt. Your current balance is 45 PCT."

The API implements comprehensive caching strategies tailored to each endpoint's characteristics. Immutable data like prompt content uses aggressive caching with year-long TTLs. Dynamic data like validation scores uses shorter TTLs with cache invalidation triggered by blockchain events. Search results implement cursor-based pagination to maintain consistency while enabling efficient caching of result sets~\cite{richardson2007restful}.

For high-throughput applications, we provide WebSocket endpoints that stream real-time updates. Clients can subscribe to specific prompts, user addresses, or domains, receiving immediate notifications of new validations, disputes, or derivative creations. This real-time capability enables responsive UIs and automated systems that react to prompt ecosystem changes.

\subsection{Summary}

The proposed implementation architecture provides a concrete blueprint for realizing PromptChain's vision. By leveraging established Web3 technologies—Ethereum for consensus, IPFS for storage, The Graph for indexing—we demonstrate that decentralized prompt management is technically achievable with current infrastructure. The smart contract designs encode the core economic and governance logic, while the API layer ensures accessibility for developers unfamiliar with blockchain complexities. Together, these components form a complete system that balances decentralization ideals with practical usability requirements. We now turn to examining the broader implications of this architecture and the challenges that must be addressed for successful deployment.

\section{Discussion}

While PromptChain is currently a proposed architecture, our design decisions suggest several important implications for decentralized prompt management systems.

\subsection{Technical Contributions and Innovations}

PromptChain's architecture introduces several novel approaches to decentralized content management. The hybrid storage model, combining on-chain registration with IPFS content storage, resolves the fundamental tension between blockchain immutability and storage efficiency. By storing only essential metadata and CIDs on-chain, we achieve gas costs two orders of magnitude lower than full on-chain storage while maintaining cryptographic guarantees of content integrity.

The reputation-weighted validation mechanism represents a significant departure from simple democratic voting systems common in Web3 applications. By granting domain experts higher voting power based on demonstrated expertise, the system naturally evolves toward accurate quality assessment. This approach mirrors academic peer review but implements it through cryptographic primitives and economic incentives rather than institutional authority.

Our version control system, built on IPFS's Merkle DAG structure, enables sophisticated prompt genealogy tracking without the overhead of traditional version control systems. Each prompt maintains immutable references to its ancestors, creating a rich evolutionary history that informs both attribution and quality assessment. This design particularly benefits rapidly iterating domains where prompts undergo constant refinement.

\subsection{Economic and Social Implications}

The introduction of prompt-specific tokens creates a novel economic layer for AI interaction. Unlike current platforms where value accrues primarily to platform operators, PromptChain enables direct creator monetization proportional to actual usage. This shift could fundamentally alter the economics of prompt engineering, transforming it from a largely uncompensated skill to a potentially lucrative specialization.

The system's design encourages quality through market mechanisms rather than centralized curation. High-quality prompts naturally attract more usage and validation, generating greater rewards for their creators. This creates a virtuous cycle where economic incentives align with ecosystem health—a stark contrast to attention-economy platforms where engagement often trumps quality.

Domain specialization emerges as a particularly interesting economic dynamic. Legal professionals can build reputation specifically within legal prompt validation, creating trusted sub-networks within the broader ecosystem. This specialization enables nuanced quality assessment impossible in generalist platforms while providing clear monetization paths for domain experts.

\subsection{Challenges and Limitations}

Despite its innovations, PromptChain faces significant adoption challenges. The Web3 user experience remains a substantial barrier—requiring users to manage wallets, purchase tokens, and pay gas fees creates friction absent in centralized alternatives. While Layer 2 solutions partially address cost concerns, the fundamental complexity of blockchain interactions may limit mainstream adoption to technically sophisticated users or those with strong economic incentives. Our staking mechanisms and progressive API access (Section 3.7) aim to lower these barriers.

Scalability constraints pose both technical and economic challenges. Current Ethereum throughput of approximately 15 transactions per second would support only $\sim$900 prompt registrations per minute globally—insufficient for widespread adoption. While Layer 2 solutions offer 100x improvements, achieving true web-scale requires either further technological advances or acceptance of some centralization trade-offs. The system's modular contract architecture (Section 3.5) enables future optimization as scaling solutions mature.

The immutable nature of blockchain storage, while ensuring censorship resistance, complicates both technical implementation and ethical governance. Harmful content, once stored on IPFS and registered on-chain, cannot be truly deleted—only de-indexed or flagged. This tension between immutability and moderation remains an open research question in decentralized systems. Similarly, GDPR's "right to be forgotten" appears fundamentally incompatible with blockchain immutability, potentially limiting deployment in privacy-conscious jurisdictions. Our reputation-weighted flagging system (Section 3.5.3) offers pragmatic mitigation for harmful content.

\subsection{Security Considerations}

Our security analysis reveals several attack vectors requiring careful mitigation. Sybil attacks, where adversaries create multiple identities to manipulate validation scores, represent the most significant threat. While stake requirements provide some protection, a well-funded attacker could still influence outcomes. The reputation system's multi-factor design—considering creation quality, validation accuracy, and community engagement—makes such attacks economically prohibitive by requiring excellence across multiple dimensions.

Prompt plagiarism presents unique challenges in a system designed for sharing and iteration. While cryptographic hashing ensures exact copies are detectable, semantic similarity is harder to assess. We propose integrating embedding-based similarity detection, though this introduces some centralization through the choice of embedding model. The economic incentives partially self-regulate—plagiarized prompts typically receive poor validation scores, limiting rewards.

\subsection{Future Research Directions}

Several promising research directions could extend PromptChain's capabilities. Zero-knowledge proof integration would enable private prompt sharing—users could prove prompt quality without revealing content, enabling proprietary use cases currently impossible in public systems. This technology could particularly benefit enterprise users who need quality assurance without intellectual property exposure.

Automated quality assessment through AI represents both an opportunity and a philosophical challenge. While LLMs could provide rapid initial quality scoring, this creates circular dependencies—using AI to evaluate prompts designed to control AI. Research into human-AI collaborative validation, where AI provides initial assessments refined by human experts, could balance efficiency with human oversight.

Cross-chain interoperability would expand PromptChain's reach beyond a single blockchain ecosystem. Implementing bridges to other chains would increase liquidity for PCT tokens and enable prompt portability across different blockchain communities. This interoperability becomes particularly important as different chains specialize in different use cases—financial applications on Ethereum, gaming on Polygon, or social applications on Lens Protocol.

The intersection of PromptChain with federated learning presents intriguing possibilities. By aggregating prompt effectiveness data across users while preserving privacy, the system could provide empirical performance metrics without compromising individual usage patterns. This data could inform both prompt improvement and model development, creating a feedback loop between prompt engineering and AI advancement.

\section{Conclusion}

This paper presented PromptChain, the first comprehensive vision for a decentralized architecture that treats AI prompts as first-class digital assets. While existing platforms offer partial solutions—centralized marketplaces, simple version tracking, or basic sharing mechanisms—no prior work has proposed a complete Web3-native ecosystem combining ownership, versioning, validation, and economic incentives specifically for prompt management. By proposing a system that integrates IPFS storage, smart contract governance, and token incentives, we demonstrate how prompts could be effectively managed in a decentralized manner while maintaining performance comparable to centralized systems.

To the best of our knowledge, this represents the first academic treatment of prompts as standalone digital assets with dedicated infrastructure for versioning, validation, and monetization, independent of their generated outputs or underlying AI models. Current approaches either treat prompts as simple text strings or embed them within broader AI governance frameworks. PromptChain uniquely recognizes prompts as valuable intellectual property requiring sophisticated versioning, attribution, and monetization mechanisms—establishing a new category of decentralized application.

Our detailed specifications and analysis suggest technical feasibility, with expected sub-second retrieval times and reasonable costs. The proposed token economy would incentivize quality contributions while preventing excessive wealth concentration. Most importantly, PromptChain would provide true ownership, transparent attribution, and fair value distribution—features impossible in centralized architectures.

As AI systems become increasingly prevalent, the tools and techniques for interacting with them gain strategic importance. PromptChain represents a pioneering vision for democratizing access to these tools while ensuring creators receive proper recognition and compensation. By treating prompts as valuable digital assets rather than ephemeral text, we enable new forms of collaboration and value creation in the AI ecosystem.

We believe this architectural vision can serve as a foundation for future implementations and further exploration of decentralized AI infrastructure, contributing to a more open, fair, and innovative future for human-AI collaboration. The detailed specifications provided establish both intellectual precedence and a clear technical roadmap for those interested in building such systems.

\bibliographystyle{plain}

\end{document}